# Charge Sharing Effect on 600 μm Pitch Pixelated CZT Detector for Imaging Applications [*]

Yin Yong-Zhi(尹永智), Liu Qi(刘奇), Xu Da-Peng(徐大鹏), Chen Xi-Meng(陈熙萌)[†]

School of Nuclear Science and Technology, Lanzhou University, Lanzhou 730000, China

## Abstract

We are currently investigating the spatial resolution of highly pixelated Cadmium Zinc Telluride (CZT) detector for imaging applications. A 20 mm ×20 mm ×5 mm CZT substrate was fabricated with 600 μm pitch pixels (500 μm anode pixels with 100 μm gap) and coplanar cathode. Charge sharing between two pixels was studied using collimated 122 keV gamma ray source. Experiments show a resolution of 125 μm FWHM for double-pixel charge sharing events when the 600 μm pixelated and 5 mm thick CZT detector biased at -1000 V. In addition, we analyzed the energy response of the 600 μm pitch pixelated CZT detector.

## 1    Introduction

The applications of Cadmium Zinc Telluride (CZT) in medical imaging applications, such as Positron Emission Tomography (PET) and Single Photon Emission Computed Tomography (SPECT), are now widely investigated due to its high energy resolution, high spatial resolution, and room temperature operability [1-4]. With the well-known small pixel effect [5], pixelated CZT detector's energy resolution and spatial resolution have been shown to improve as the pixel size decreases. This effect was proved for the CZT detectors with larger pixel size (typically > 1 mm) [1, 6]. However, experiments and Monte Carlo simulations have shown this tendency may not hold true when the pixel size becomes very small relative to the size of charge cloud created by the gamma ray interacting within the CZT detector [7-9]. Under this condition, charge sharing events

[*] Project supported in part by NSFC (11305083) and in part by the Fundamental Research Funds for the Central Universities (lzujbky-2013-5).
[†] E-mail:chenxm@lzu.edu.cn

among multiple pixels appear and decrease the spatial resolution of the imaging detector.

In the imaging applications of pixelated CZT detector, the image resolution was limited by the intrinsic spatial resolution of that detector. Commonly, only single-pixel photopeak events were used in the image reconstruction. The charge sharing events in the image reconstruction was not included even this part events are dominant for the highly pixelated CZT detector [7]. One of the main challenges is the interpolation algorithm for charge sharing events are not very clear. To overcome this problem, the charge sharing range of the pixelated CZT detector should be finely measured.

To better understand the small pixel effect and charge sharing in the pixelated CZT detector with sub-millimeter pixel size, intensive studies were developed and reported. In the past researches, the sub-millimeter pixel size of CZT detector included 0.25 mm [10], 0.4 mm [11], and 0.6 mm [12], 0.8 mm [13], etc. Experiments show there are several key factors can contribute to charge sharing, including the electron-hole cloud size, the diffusion of charge carriers, Compton scattering events, and Characteristic X-rays. In this work, we focused on the effects of pixel size and charge diffusion. CZT detectors with 600 μm pitch (500 μm pixel and 100 μm gap) and 5mm thick were fabricated and characterized. We studied the charge sharing characteristics and energy response of pixelated CZT detector to evaluate the feasibility of using such kind of detectors for high resolution PET/SPECT imaging applications.

## 2     Experimental Setup

### 2.1    CZT Detector

A 20 mm × 20 mm × 5 mm Modified High-Pressure Bridgman (MHB) CZT substrate was contacted metal pixels in a class-100 clean room. The CZT substrate was first polished and etched with a 5%-95% Br-Methanol solution to improve both the electrical property and the adhesion of the contacts [14]. Standard photolithographic process and electron beam evaporator were used to deposit pixel contacts. The finished CZT coplanar cathode surface is made of 125 nm thickness Gold, which is high-work function material that can reduce leakage current and improve energy resolution. The anode pixel surface is deposited with 100 nm Titanium, which has a relative low-work function and has been shown to provide the best energy resolution among four anode materials investigated: Indium, Titanium, Chromium and Gold [15].

Fig.1 (Left) shows a finished CZT detector, which has two kinds custom designed pixel patterns. The 9-pixel part was used to evaluate the charge sharing effect in current experiments. At the center of this 9-pixel pattern, there are nine 500 μm ×500 μm anode pixels with 100 μm gaps. The 8 neighboring anode pixels are connected to eight circular readout pads that are 1.4mm in diameter and 2.5 mm away. This layout allowed us to read out 9 anode signals using pogo pins between 2.5 mm distance instead of 0.6 mm tight space. During the experiment, the CZT detector was biased at -1000 V. We used a collimated Co-57 source scanning across two neighbor pixels, pixel A4 and pixel A5. To validate such a special design, we simulated the weighting potentials and electric field of this 9-pixel pattern CZT detector based on a solution of the 3D Laplace equation [8]. Simulation shows the central anode pixel (Anode A5) in this special 9-pixel geometry is a good approximation of the regular 600 μm pitch pixelated CZT detector that we intend to evaluate.

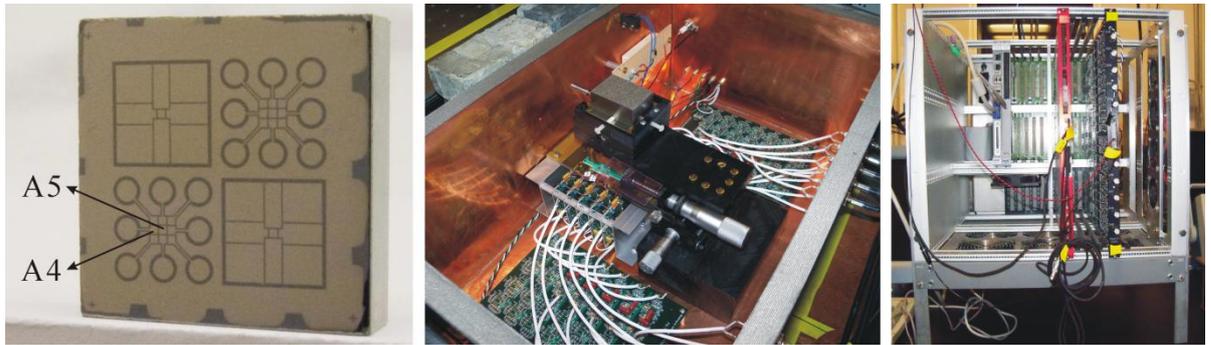

Fig. 1    (Left) Finished 20 mm ×20 mm ×5 mm CZT detector with spatial 9-pixel pattern; (Middle) Custom designed readout electronics and the collimator; (Right) VME crate with a 500MHz flash ADC board.

## 2.2    Data Acquisition System

The readout electronics of this pixelated CZT detector was shown in Fig.1 (Middle). The nine anode signals of the above CZT detector were read out through 9 spring-loaded pogo pins. The pogo pins were arranged to form a 3×3 array spaced at 2.5 mm pitches and connected with the nine anode pixels. All anode and cathode signals were read out through low noise preamplifiers (A500, Amptek, Bedford, MA) and shaping boards. The CZT detector and its readout boards are mounted inside a copper box to minimize noise pickup. A 2-D translation stage is mounted next to the CZT detector (also inside the copper enclosure) to hold a radioactive source and a collimator in order to provide a collimated gamma ray beam for spatial resolution and charge sharing measurements. The minimal step size of the translation stage is 10 μm.

A collimated Co-57 radioactive source was used to scan across three pixels. A custom holder was made to mount collimator and gamma ray source to the 2-D translation-stage. The collimator was made of four tungsten blocks to create two orthogonal slit collimators that form a size adjustable beam source. During the experiment, the entrance of the collimator keeps 50 μm, and the effective collimated beam size is 160 μm × 160 μm. This small collimator size makes the radiation flux relative low. In the data acquisition, it takes one day to accumulate an energy spectrum.

Amplified and shaped by the preamplifiers and the shaping circuits, signals were digitized by a typical DAQ system, which includes a 10-channel Flash ADC, a clock/trigger Board and VME bus, as shown in Fig.1 (Right). The flash ADC has a maximal sampling rate of 500 MHz and 8-bit depth. Each channel has on-board constant fraction discriminator (CFD) to set the energy threshold. If the signal on any electrode exceeded its threshold, the system was triggered and waveforms of all nine anode channels and cathode channel were measured and recorded for off-line analysis. In the experiment, we use 10 ns sampling rate to reduce the data size.

## 3      Results and Discussions

### 3.1    Charge Sharing

Charge sharing effect between the central pixel and its left neighbor pixel was measured using collimated Co-57 beam. To evaluate the charge sharing range, we scanned across two neighboring pixels. The collimated beam has an effective size of 160 μm × 160 μm on the CZT cathode surface. During the experiments, we moved the collimator in a step size of 50 μm.

Fig.2 shows scatter plots of charge signals from 2 neighboring anode pixels (central pixel A5 and its left neighbor pixel A4) at different source locations as a collimated 122 keV gamma ray beam was stepped from one pixel to the other. We only show six positions in Fig.2 to illustrate the changes of charge sharing between the two pixels. From the scatter plots, we can clearly see that more charge sharing events occur when the collimated beam hits the gap between the two pixels (position '240 μm' and '260 μm'). We could see both charge sharing events and single-pixel events in the scatter plots of position '150μm' and '350μm'. Only single-pixel events could be found when the collimated beam hits the middle of anode pixels (position '0μm' and '500μm').

Fig.2 proves that the number of double-pixel charge sharing events reaches its peak value when the collimated gamma ray beam hits the center of the gap between adjacent anode pixels.

The number of double-pixel charge sharing events approaches zero when the collimated gamma ray beam hits the center of anode pixels. These results suggest a potential way to estimate the charge cloud dimension in CZT experimentally by measuring count profile of double-pixel charge sharing events when collimated gamma ray beam is stepped across 600 μm pitch pixelated CZT detector.

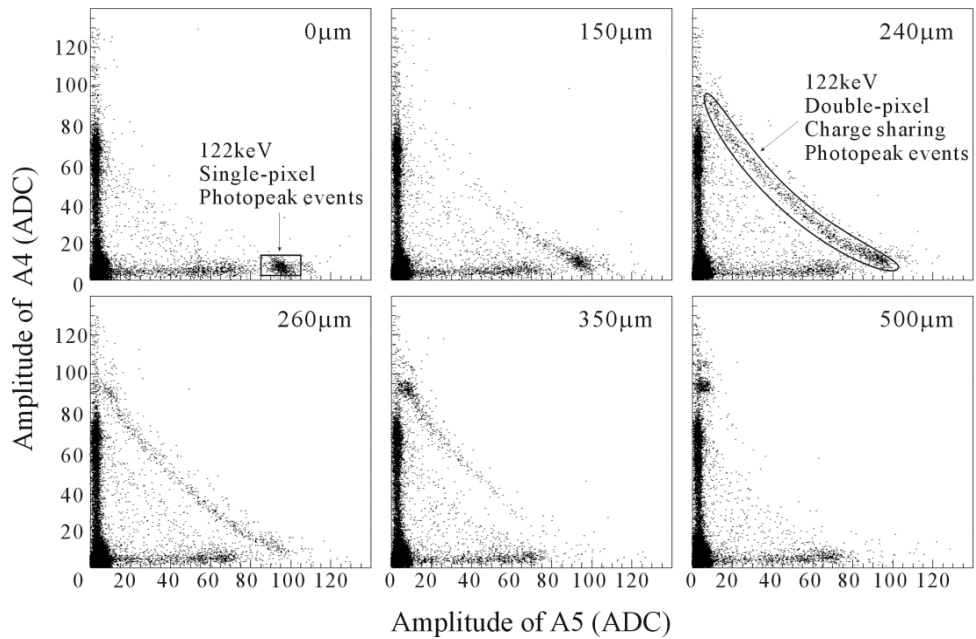

Fig. 2  Charge sharing between two neighbor pixels, A5 and A4, when collimated 122 keV beam hits different locations in the 600 μm pitch pixelated CZT detector measurements.

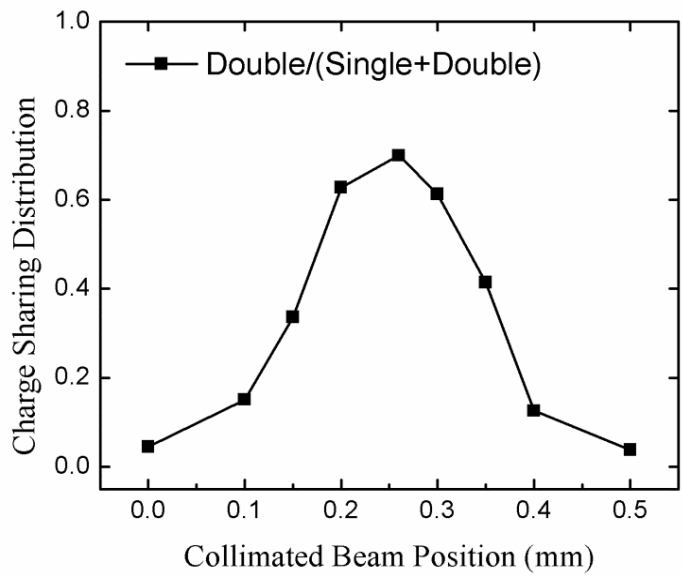

Fig. 3  Charge sharing distribution of 600 μm pixelated CZT when the collimated 122 keV beam scans across two pixels. This profile shows the ratio of double-pixel photopeak events versus single-pixel photopeak events plus double-pixel photopeak events.

In addition, we found that the line of 122 keV photopeak events bended in the scatter plots of position '240 μm' and '260 μm'. This phenomenon indicates that the pixelated CZT detector we used suffers charge loss on the gap between two pixels. For photopeak events of double-pixel charge sharing, the charge will finally collect by two pixels when charge cloud drifts and arrives. But the total charge collected by two pixels is smaller than the charge collected by single-pixel photopeak events. This was also proved by the energy spectrum in Fig.4 and Fig.5.

Fig.3 shows the count profile of double-pixel photopeak events measured by the central pixel A5 and its left neighboring pixel A4 when collimated 122keV gamma ray was stepped across the 500 μm anode and the 100 μm gap. Results show that for a Co-57 source (122keV), the FWHM of count profiles of double-pixel events is 125 μm after the dimension of the collimated gamma ray beam is subtracted from the measured count profiles in Fig.3. This value indicates that, for our 600 μm pixelated CZT detector, the charge sharing range is bigger than the gap size but not significant in the 122 keV beam measurement.

## 3.2 Energy Response

Fig.4 (Left) shows the energy spectrum of 600 μm pixelated CZT detector when the collimated Co-57 beam hits the central pixel of 9-pixel pattern, corresponds to the position '0 μm' in Fig.2. Both 122 keV photopeak and 136 keV photopeak are clearly seen in the energy spectrum. For 122 keV events, the energy resolution of central pixel is 6.8%. Meanwhile the energy resolution of the 1.4 mm readout pad shows 5.2%, if we move the collimated beam to hit that readout pad. The electronic noise contribution is about 3 keV, which is not subtracted in the energy resolution. This suggests that for finely pixelated CZT detectors, such as 600 μm pixelated detector used in our experiments, the benefits from small pixels effect that have been widely reported may start to diminish because small pixels might could not fully collect the charge. The charge collection by small pixels would be even worse for higher energy photons such as 511 keV gamma rays that produce larger charge cloud.

Fig.4 (Right) shows the scatter plot of charge signal amplitudes of central anode and cathode, when the collimated beam hits central pixel. The 122 keV photopeak events clearly show the depth dependence. Most of the 122 keV photopeak events are collected near the top of the detector (close to the cathode side). The cathode signal amplitude has depth of interaction (DOI) information because the holes collected by cathode are decreasing with the travel distance of holes. Based on this phenomenon, the DOI correction of energy resolution for Co-57 beam was not included in this paper.

Fig.5 (Left) shows the energy spectrum of double-pixel charge sharing events when the collimated Co-57 beam hits the gap between two pixels, corresponds to the position '240 μm' in Fig.2. We plot the summed signal amplitude of central pixel A5 and the left neighboring pixel A4. The double-pixel photopeak events of 122 keV show an 8% energy resolution. Fig.5 (Right) shows the scatter plot of charge signal amplitudes of central anode and cathode. Compare the Fig.4 and Fig.5, both 122 keV photopeak and 136 keV photopeak could be seen in the energy spectrum and the scatter plot. But the energy resolution of single-pixel photopeak events is better than the energy resolution of the double-pixel charge sharing photopeak events for 600 μm pixelated CZT detector.

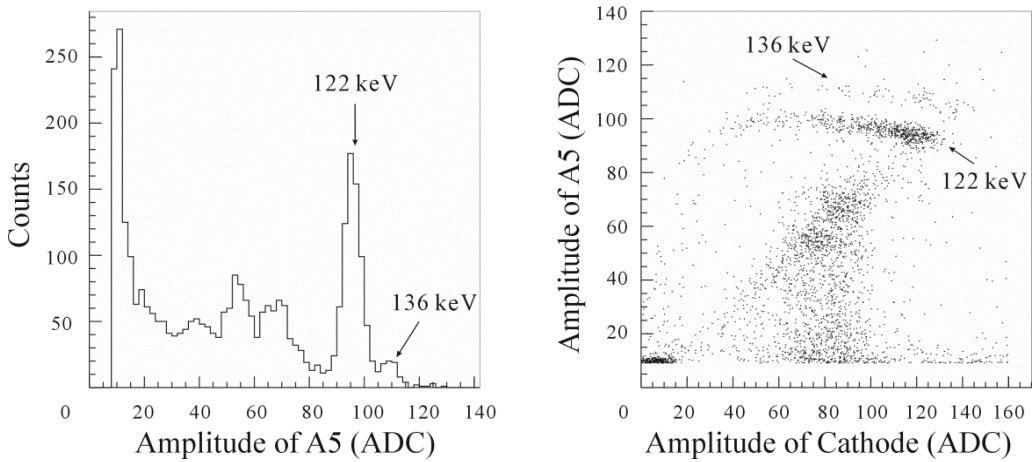

Fig. 4  Energy spectrum (Left) and scatter plot of anode versus cathode (Right) when collimated Co-57 beam hits the central 600 μm pixel of the CZT detector.

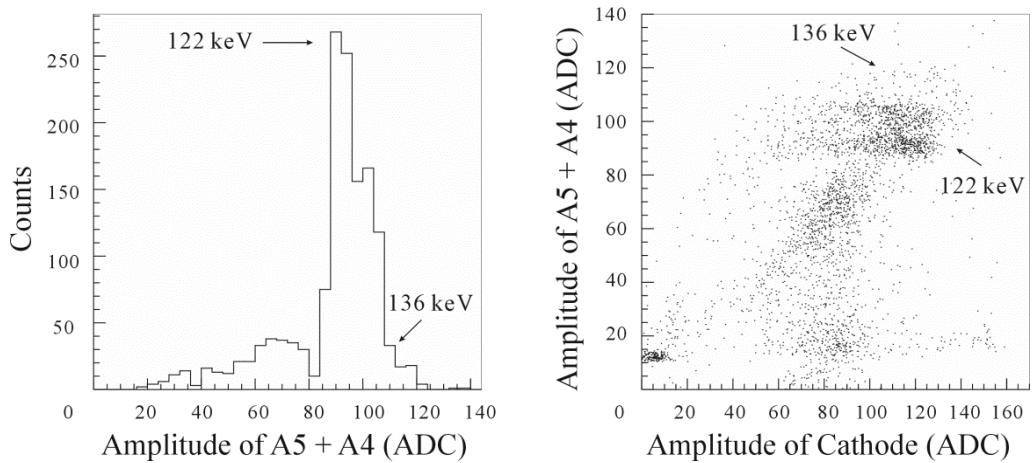

Fig. 5  Energy spectrum of summed signal amplitude of two neighbor pixels (Left) and scatter plot of summed anode versus cathode (Right) when collimated Co-57 beam hit the gap between two neighbor pixels of 600 μm CZT detector.

## 4    Conclusions

We have studied charge sharing and energy response of 600 μm pitch pixelated CZT detector by a collimated 122 keV gamma-ray beam. The experiment clearly shows more charge sharing events occur when the collimated beam hits the gap between the two pixels. The number of double-pixel charge sharing events approaches zero when the collimated beam located at the center of the anode pixel. The analysis of scatter plots between two neighboring pixels suggests a resolution of 125 μm FWHM for double-pixel charge sharing events when the 600 μm pixelated and 5mm thick CZT detector biased at -1000 V. And experiments proved the energy resolution of single-pixel photopeak events is better than the energy resolution of the double-pixel charge sharing photopeak events for 600 μm pixelated CZT detector.

## Acknowledgment

The authors thank Dr. Yuan-Chuan Tai, Dr. Heyu Wu, Dr. Sergey Komarov and Dr. Henric Krawczynski at Washington University in St. Louis for their support and valuable discussions.